\documentstyle[epsfig,subfigure,12pt]{article}
\begin{document}
\newcommand {\ber} {\begin{eqnarray*}}
\newcommand {\eer} {\end{eqnarray*}}
\newcommand {\bea} {\begin{eqnarray}}
\newcommand {\eea} {\end{eqnarray}}
\newcommand {\beq} {\begin{equation}}
\newcommand {\eeq} {\end{equation}}
\newcommand {\state} [1] {\mid \! \! {#1} \rangleg}
\newcommand {\sym} {$SY\! M_2\ $}
\newcommand {\eqref} [1] {(\ref {#1})}
\newcommand{\preprint}[1]{\begin{table}[t] 
           \begin{flushright}               
           \begin{large}{#1}\end{large}     
           \end{flushright}                 
           \end{table}}                     
\def\Acknowledgements{\bigskip  \bigskip {\begin{center} \begin{large}
             \bf ACKNOWLEDGMENTS \end{large}\end{center}}}

\newcommand{\half} {{1\over {\sqrt2}}}
\newcommand{\dx} {\partial _1}

\def\Dslash{\not{\hbox{\kern-4pt $D$}}}
\def\cmp#1{{\it Comm. Math. Phys.} {\bf #1}}
\def\cqg#1{{\it Class. Quantum Grav.} {\bf #1}}
\def\pl#1{{\it Phys. Lett.} {\bf #1B}}
\def\prl#1{{\it Phys. Rev. Lett.} {\bf #1}}
\def\prd#1{{\it Phys. Rev.} {\bf D#1}}
\def\prr#1{{\it Phys. Rev.} {\bf #1}}
\def\np#1{{\it Nucl. Phys.} {\bf B#1}}
\def\ncim#1{{\it Nuovo Cimento} {\bf #1}}
\def\lnc#1{{\it Lett. Nuovo Cim.} {\bf #1}}
\def\jmath#1{{\it J. Math. Phys.} {\bf #1}}
\def\mpl#1{{\it Mod. Phys. Lett.}{\bf A#1}}
\def\jmp#1{{\it J. Mod. Phys.}{\bf A#1}}
\def\aop#1{{\it Ann. Phys.} {\bf #1}}
\def\mycomm#1{\hfill\break{\tt #1}\hfill\break}

\begin{titlepage}
\titlepage
\rightline{TAUP-2412-97}
\rightline{\today}
\vskip 1cm
\centerline{{\Large \bf Screening
and confinement  in large $N_f$
    $QCD_2$}}
\centerline{{\Large \bf and in  $N=1$ \sym}}
\vskip 1cm

\centerline{A. Armoni and  J. Sonnenschein
\footnote{Work supported in part by the Israel Science Foundation,
and  the US-Israel Binational
Science Foundation }}

\vskip 1cm
\begin{center}
\em School of Physics and Astronomy
\\Beverly and Raymond Sackler Faculty of Exact Sciences
\\Tel Aviv University, Ramat Aviv, 69978, Israel
\end{center}
\vskip 1cm
\begin{abstract}
The screening nature of the potential between external quarks
in massless $SU(N_c)$ $QCD_2$ is derived using an expansion
in $N_f$- the number of flavors.
Applying the same method to the massive model, we find
a confining potential.
We consider the  $N=1$ super Yang Mills theory, reveal certain
problematic aspects of  its bosonized version and show the
associated screening behavior by applying a point splitting
method to the scalar current.
\end{abstract}
\end{titlepage}
\newpage

\section{Introduction}
Large $N_c$ expansion\cite{tHooft}, and strong coupling expansion
\cite{FS}
were shown to be powerful tools in analyzing various aspects of
$QCD_2$. In the present work we demonstrate the usefulness
of yet another technique, that of expanding  in
large number of flavors $N_f$. We apply this method
to derive further evidence for the screening behavior
between external quarks in the massless theory,  and the confining
one in the massive theory\cite{gross}.

 The  screening nature
of the potential  was argued in ref.\cite{gross}
by substituting a static  (abelian) solution of the equations of
motion for the gauge configuration  into the expression of the
potential.
This approach falls short in proving the screening nature
of the interaction between the external quarks because of the
following drawbacks. (i) Classical configurations
cannot always recast for the quantum behavior of a system.
 (ii) Treating the external quarks as classical
 sources is strictly speaking justified
only for large color representations in a similar way
that only particles with large angular momentum can be described in
quantum mechanics as c-number quantities.

In the present paper we suggest
 improvements that enable one to overcome these obstacles.
To justify the fact that functional integral is dominated by the
saddle-point configurations we introduce $N_f$ flavor degrees of
freedom and  analyze the system in the limit of large  $N_f$. It
is easy to realize, especially in the bosonization formulation,
 that the action can be brought into a form where it is multiplied
by an overall factor of $N_f$.  This obviously implies that
 ${1\over N_f}$ plays the role of
$\hbar$ so that  the limit of $N_f\rightarrow \infty$
corresponds to the classical limit of  $\hbar \rightarrow 0$.

Moreover, the abelian gauge configuration that solves the equations
of motion and is used in \cite{gross} to derive a screening
potential, can be used as a first iteration in
 a systematic expansion of the gauge configuration
in powers of ${1\over N_f}$. One can thus check whether the
screening nature persists also in corrections which are
proportional to  higher powers of  ${1\over N_f}$.
However, if one presents the external
quarks  in terms of commuting functions as was used in \cite{gross},
then the  iterated gauge field solution at any order will remain
abelian if  the leading one is.
By abelian we mean that commutator terms in the equations of
motions vanish. One may  be suspicious that in that way one
abelianize $QCD_2$ in an unjustified manner and hence
the form of the  potential cannot  be faithfully
examined.
This situation can be avoided either by searching genuine
non-abelian solutions of the equations of motions \cite{FScs}
or by using a different method of representing the external
quarks so that the non-abelian commutator terms do not vanish.
A proposal of this nature was introduced in refs.\cite{adler,giles} where
the density of the external quarks was represented in terms of
non-commuting matrices.
We compute the potential including corrections of next and next
to next order corrections and show that it tends to a constant
for large separation distance, thus it exhibits a screening
behavior.

Two dimensional massless $QCD$ admits a universal
nature\cite{kutasov2}. The introduction of mass terms to the
quarks creates a  much more dramatic alteration in the system than
in the real world 4D $QCD$. This property manifests itself in
the baryonic spectrum\cite{FS} as well as in the form of the
potential between external quarks. External quarks in the
fundamental representations that were screened by massless adjoint
dynamical fermions, find themselves in the confining phase once
the dynamical fermions acquire mass.
In ref.\cite{gross} this phenomenon was derived
for the analog abelian case as well as for $QCD_2$ with
two and three colors. In \cite{FScs} it was further shown for
a general $SU(N_c)$ gauge group by bosonizing both the dynamical
and the external quarks and the conditions for finite energy quark
configurations were analyzed.
An explicit determination of the potential, in a similar manner
to way it was done in the massless case, is technically much more
tedious  due to the form of the bosonized mass term.
Once again the expansion in large $N_f$ comes to our rescue.
The mass term can be expanded in terms of a
(non-local) power series of  the  color current divided by $N_f$.
Restricting the expansion  to the lowest power of ${1\over N_f}$
renders the equations of motion in terms of the currents to
tractable ones. Using this method  we show that
the leading solution for the potential exhibits
a confining behavior.

A natural framework that ``embeds" the  $QCD_2$
model with adjoint \newline  fermions is that of $N=1$ super Yang-Mills
theory. In ref. \cite{gross} the potential of that model is
  conjectured to be a screening one.
We  show that a naive bosonization of the fermionic degrees of
freedom of the model leads to a wrong picture. It is further
shown that by considering the contribution of the scalar field
 to the one loop vacuum polarization one indeed discover the
non-confining potential.

 The paper is organized as follows. In section 2  the
equations of motion  of  multi-flavor massless  $QCD_2$  in the
presence of external quarks are derived in the $A_-=0$ gauge. A
large $N_f$
 iteration analysis is described in section 3.  The potential
is shown to be a screening one even for a non-commuting
presentation of the external density.
Section 4 is devoted to the massive  multi-flavor case.
Expressing the mass term in a power series in the currents we
derive the confining potential for that  model.
An analysis of the potential for the $N=1$ supersymmetric
Yang-Mills theory is presented in section 5. We show that a one
loop effect in the scalar sector which is the analog of the
anomaly term for the fermions combines with the latter to produce
a screening potential.

\section{Multi-flavor $QCD_2$ with external source}
Massless multi-flavor $QCD_2$ with fermions in the fundamental
representation of $SU(N_c)$ and external current coupled to the gauge field
 is described by the following action
\beq
\label{qcd} S=\int d^2x \ tr (-{1\over 4}F_{\mu\nu}^2+i\bar \Psi\Dslash\Psi -e
 A_\mu j^\mu_{ext})
\eeq
where $\Psi = \Psi ^{i\alpha}$, $i=1 \dots N_c$ and $\alpha =1 \dots
N_f$ and the trace is over both the color and flavor indices.
Variation with respect to the Dirac field and gauge fields leads to the
 following
equations of motion:
\bea
 && \label{vector}    D_\mu J^\mu =0 \\
 && \label{external} D_\mu j^{\mu}_{ext}=0 \\
 && \label{gauge}     D_\mu F^{\mu\nu}=e(J^\nu + j^\nu_{ext})
\eea
where $D_\mu = \partial_\mu + e[A_\mu,\ ]$.
 The first \eqref{vector} and the second \eqref{external} equations
 are  the covariant conservation of the dynamical  and external
vector currents,
 respectively.
In addition the chiral anomaly equation reads
\beq
\label{axial}       D_\mu J^{5\mu} = {eN_f\over 2\pi} \epsilon^{\mu
   \nu} F_{\mu\nu} \\
\eeq
where $J^{5\mu} = \bar \Psi \gamma ^ 5 \gamma ^\mu \Psi$.
 The anomaly equation can be obtained by the equations of
motion of the bosonized action of \eqref{qcd} (see for instance\cite{FS}).

 In order to maintain gauge
invariance of  the action
\eqref{qcd}, the external current should be covariantly conserved.
This implies that, in contrast to the abelian case, one cannot set $j^0_{ext}$
 as a
 time
independent function and then  fix $j^1_{ext}$ to be zero.
Instead  we  fix the value of $j^0_{ext}$ and treat $j^1_{ext}$
as a ``dynamical" variable of the problem.

By choosing the $A_-=0$ gauge, using    light cone coordinates
$x^\pm={1\over {\sqrt 2}}(x^0\pm x^1)$, and
  the two dimensional property
$J^{5\mu}=-\epsilon ^{\mu\nu}J_{\nu}$,  we translate equations
\eqref{vector},\eqref{axial} and \eqref{gauge} into
\bea
\label{seta}
 && \partial _+ J^+ + \partial _- J^- +e[A_+,J^+]=0 \\
 && \partial _+ J^+ - \partial _- J^- +e[A_+,J^+]= {eN_f \over
   \pi}\partial _- A_+ \nonumber \\
 && -\partial _-^{\ 2}  A_+ =e(J^+ + j^+_{ext}) \nonumber
\eea

In our analysis we will be  interested in  static solutions of the
equations that correspond to static  external sources. For this
 case the set of equations combined with the external source
constraint takes the following simplified  form
\bea
\label{set}
 && \half \partial_1 J^+ + e[A_+,J^+]=-{eN_f\over 2\pi}\half \partial_1 A_+ \\
 && {1\over 2} \partial_1^2 A_+ = -e(J^+ +\half \rho +\half j) \nonumber \\
 && \partial_1j +\half e [A_+,\rho]+\half e[A_+,j]=0 \nonumber
\eea
where we have used the notation $j \equiv j^1_{ext}$ and  $\rho \equiv j^0_{ext}
 $.
 Given $\rho$, equations  \eqref{set}   can be used
 to determine the dynamical variables $A_+$, $ J^+$ and $j$ once
 boundary conditions are specified.

The gauge field  $A_+$ itself, is not the physical quantity of
interest, however, once it is determined, the potential energy between the
external charges can be found by substituting  it  in the effective
action.  Using the equations of motion the potential takes
the following form:
\beq
  \label{presc} V={1\over 2} e \int dx\ tr (A_+ j^+_{ext}) = {1\over 2}e \int
dx A^a_+ \half (\rho+j)^a
\eeq
where $a$ are the adjoint indices  of the color group $a=1,...,N_c^2-1$.
As was discussed in the introduction, the
 behavior of the potential as a function of the distance between the external
 quark and the anti-quark determines whether the theory  is
 screening or confining. A linear potential means a constant force, namely,
  confinement, while a constant potential (at
 large distances) means screening.

We are now facing the question of how to incorporate the external sources.
Consider a system of a quark in the fundamental representation placed at a
distance  of $2R$ from an anti-quark that transforms in the anti-fundamental
representation. This can be expressed as the following classical c-number
function
\beq
  \label{rho} \rho ^a=\delta ^{a1} (\delta (x-R) - \delta (x+R))
\eeq
Strictly speaking, one is allowed to introduce classical charges
and neglect  quantum fluctuations  only if the external charges
 transform in  a large color representation.
(This is an analog of
the statement that only for quantities of large angular  momentum quantum
fluctuations are suppressed.)
Moreover, by choosing \eqref{rho}, there is an obvious   ``abelian''
self-consistent solution\cite{gross} of the equations \eqref{set}for which
  all the dynamical quantities ($A_+$,
 $J^+$ and $j$) points in the '1' direction and thus all the
 commutators vanish. This solution
  does not reflect the non-abelian nature of the
theory.
 This situation is not avoided also when one uses an iterative
expansion in large $N_f$, as is described in the next section.
Hence, it seems  that deriving conclusions from the abelian solution  based on
the use of \eqref{rho} is  unjustified  and may miss the true nature of
the interaction.
One way to overcome these obstacles is to search for ``truly" non-abelian
solutions of the equations. This approach was followed in ref. \cite{FScs}.
Here we proceed by  implementing   Adler's\cite{adler}
semi-classical approach  for introducing  static external quark charges. In this
approach the quarks color charges satisfy non-abelian $SU(N_c)$ color algebra
so that  the external quark charge density  takes the form
\beq
 \label{rho2} \rho ^a = Q^a \delta (x-R) + \bar Q^a \delta (x+R)
\eeq
 $Q^a$ and $\bar Q^a$ are
  in   $({\bf N_c}, 1)$ and   $(1,{\bf\bar  N_c})$ representations of
 $SU(N_C) \otimes SU(N_C)$ group respectively.
The algebra of those operators which was worked out in \cite{adler,giles}
is reviewed briefly in the appendix.
Note that unlike what followed  from the  classical expression
 \eqref{rho}, now since
 $Q^a$ and $\bar Q^a$  do not commute,
``non-abelian'' solutions are expected.
In the following section we apply a systematic
${1\over {N_f}}$ expansion of
 $A_+$ and derive such a solution.

\section {Large $N_f$ expansion and non-Abelian \newline solutions}

The form of  the set of equations \eqref{set} suggests a natural ${1\over N_f}$
 expansion.
Large $N_f$, with ${{e^2N_f}\over \pi}  \equiv  \mu^2$ fixed, means 
weak coupling  constant $e$. In this case a  procedure
of extracting a solution by
iterations can be applied in the following way.
 A   solution for  $A_+$,  $J^+$ and $j$
of the equations  expanded to a given  order in $e$
 is inserted back to \eqref{set} as a source to
determine the next order solution.
A similar treatment in four dimensions is given
 in refs.\cite{jackiw,arodz}.

 The formal expansion in $e$ is as follows
\bea
 && A_+=e  A^{(1)} + e^3  A^{(3)} + e^5 A^{(5)} + \dots \nonumber \\
 && J^+ = J^{(0)}+ e^2 J^{(2)}+ e^4 J^{(4)} +\dots \label{expansion} \\
 && j= e^2 j^{(2)}+ e^4 j^{(4)}+ e^6 j^{(6)}+ \dots \nonumber
\eea
Substituting $A_+$, $J^+$ and $j$ back in \eqref{set} one gets
equations for $A^{(i)}$, $J^{(i-1)}$
and $j^{(i+1)}$ which to lowest order in ${1\over N_f}$ take the form
\ber
  \half \dx J^{(0)} = -{1\over 2} \mu ^2 \half \dx A^{(1)} \\
  {1\over 2} \dx ^2 A^{(1)} = -(J^{(0)} + \half \rho)
\eer
Assuming vanishing currents at infinity the equations can
be rewritten as
\ber
 && J^{(0)}=-{1\over 2} \mu ^2 A^{(1)} \\
 && (\dx ^2-\mu ^2) A^{(1)} = -\sqrt 2 \rho
\eer
The solution for $A^{(1)}$ is
\beq
A^{(1)}=
 {\sqrt 2 \over {2 \mu}} (Q e^ {-\mu \mid x-R \mid} + \bar Q e^ {-\mu \mid x+R
 \mid}  )
\eeq

This is the ``abelian'' solution of \cite{gross} with a replacement of the
c-number charges with the non-commuting ones.
Whereas in the case of $QED_2$ coupled to external sources,
this is (with c-number charges) an   {\em exact}
solution, in the present case
it is the leading (in ${1\over N_f}$) contribution to $A_+$.
The next to leading order set of equations is
\ber
 && \half \dx J^{(2)} + [A^{(1)},J^{(0)}] = -{1\over 2} \mu ^2 \half \dx A^{(3)}
 \\
 && {1\over 2} \dx ^2 A^{(3)} = -(J^{(2)} + \half j^{(2)}) \\
 && \dx j^{(2)} + \half [A^{(1)},\rho] =0
\eer

Substituting $A^{(1)}$ and $ J^{(0)}$ we find
\ber
 && J^{(2)} = -{1\over 2}\mu ^2 A^{(3)} \\
  && (\dx ^2 - \mu ^2) A^{(3)} = - \sqrt 2 j^{(2)}
\eer

with $j^{(2)}$, which is determined by the previous iteration, of the form
\beq
 j^{(2)} = -\half {1\over \dx} [A^{(1)},\rho]
 = {1\over {4 \mu}} [\bar Q,Q] e^{-2 \mu R} (\epsilon (x+R) - \epsilon (x-R))
\eeq

 where $\epsilon (x)$ is the step function
  $\epsilon (x)= 1$ for $x>0$ and  $\epsilon (x)=-1$ otherwise and
 ${1\over \dx}$ denotes the integral $\int ^{x} _{-\infty} dx'$ .

The latter expression acts as a source to $A^{(3)}$ which leads
\beq
 A^{(3)}=
  {\sqrt 2 \over {4\mu ^3}} [\bar Q ,Q] e^{-2 \mu R}
  (\epsilon (x+R)(1-e^{-\mu \mid x+R \mid})
  -\epsilon (x-R)(1-e^{-\mu \mid x-R \mid}))
\eeq

Far from the sources $A^{(3)}$ approaches zero. 
 This indicates  that the potential is approaching a
constant value when one  quark is  taken to be far  from the other.
However, as we shall see,  $A^{(5)}$ will be needed to observe  the first
correction  to the potential. This calculation is written in the Appendix.

We substitute now the expression found for
 $A_+$ and $j^+$ in \eqref{presc}  to derive the following  potential
\ber
  \lefteqn {V = {1\over 2}e \int dx A_+ \half (\rho +j) } \\ &&
  = {e \over {2\sqrt 2}} \int dx (e A^{(1)}+e^3 A^{(3)} + e^5 A^{(5)}
  + \dots)(\rho + e^2 j^{(2)} + e^4 j^{(4)} + \dots) \\ &&
  = {1 \over {2\sqrt 2}} \int  dx
   \left ( e^2 A^{(1)} \rho  + e^4 (A^{(1)} j^{(2)} + A^{(3)} \rho )
            + e^6 ( A^{(1)} j^{(4)} + A^{(3)} j^{(2)} + A^{(5)} \rho )
             + \dots \right )
\eer

 All the $e^4$ terms vanish because they contain $[Q,\bar Q]Q \sim
 f^{abc} Q^a \bar Q^b Q^c $ part which vanishes. Therefore the first
 non-trivial correction is $O(e^6)$.

\bea
  \lefteqn{V(2R)=
  {e^2 \over {4\mu}} (QQ + \bar Q \bar Q + 2Q \bar Q e^{-2\mu R})}
  \label{potential} \\ &&
 +{e^6 \over {8 \mu ^5}} {[\bar Q,Q]} ^2  \left [
   {(1-e^{-2\mu R})} ^2 - 2\mu R \ e^{-2 \mu R} (1-e^{-2 \mu R})
 \right ]
  \nonumber
\eea

Let us compute the group constants that appear in \eqref{potential} .
 $QQ$ stands for $Q^a Q^a$ which is the second Casimir
 operator of the fundamental \newline
representation- $Q^a Q^a = {{N_c^2-1} \over {2N_c}}$.
  Similarly $\bar Q \bar Q = {{N_c^2-1}\over {2N_c}}$. The value of
  $Q \bar Q$ is determined easily by using $2 Q^a \bar Q ^a = (Q^a + \bar
  Q ^a)(Q^a + \bar Q^a) -  Q^a Q^a - \bar Q^a \bar Q ^a$. In the
  singlet coupled state $Q^a + \bar Q^a = 0 $ and hence $Q^a \bar Q ^a
  = -{{N_c^2-1}\over {2N_c}}$.

The value of  ${[Q,\bar Q]}^2 $ is computed by the $Q \bar Q$ algebra
(see the Appendix), it is ${[Q, \bar Q] }^2 = {{N_c^2}\over
  2}{{N_c^2-1}\over {2N_c}}$.

Defining $d \equiv 2R$ , the potential takes the form:
\bea
  \lefteqn{V(d)=
 \mu {\pi \over {2N_f}} {{N_c^2-1}\over {2N_c}} (1-e^{-\mu d}) }
\label{potential2} \\ &&
 +\mu {({\pi \over {2N_f}})}^3 {{N_c^2}\over 2}{{N_c^2-1}\over {2N_c}} \left (
   {(1-e^{-\mu d})} ^2 - \mu d \ e^{- \mu d} (1-e^{- \mu d})
 \right )
  \nonumber
\eea

Thus, the potential that includes
the first correction to the abelian one approaches a constant value
at large distances where the force between the  external quark and the
anti-quark vanishes.
The obvious question now is whether we can infer from
this result that the screening nature of the interaction remains valid
to all orders in ${1\over N_f}$. We cannot provide a general proof
of that statement, but we believe that indeed that is the exact
nature of the interaction.  This is based on the following argument.
 The structure
of all higher contributions is $(\dx ^2 - \mu ^2) A = - j $, where $j$
is determined by previous iterations. Therefore, $j$ would always
vanish far from the sources and consequently the gauge field $A$ would exhibit
 the
same behavior.

\section{Large $N_f$ expansion of  massive  $QCD_2$}
Let us consider now the case of massive dynamical quarks. Whereas, for the
massless case   the equations that determine the potential
 can be derived  using both  the fermionic
picture and the bosonized one, here we can apply our analysis only in the
bosonization description.
 The bosonized action
 of massive $QCD_2$ with $N_f$ fundamental representations
in the gauge  $A_-=0$ takes the following form\cite{FS}

\ber \label{ghlboz}
 \lefteqn{S=S^{WZW}_{(N_c)} (g) + S^{WZW}_{(N_f)} (h)
 +{1\over 2} \int d^2x \
 \partial _\mu \phi \partial ^\mu  \phi + S^{WZW}_{(1)} (l) + }\\ &&
 \int d^2x \ tr \left [m^2:(ghl \exp ({-i\sqrt {{4\pi} \over
       {N_cN_f} }}\phi) +\exp ({i\sqrt
  {{4\pi}\over {N_cN_f}}}\phi) l^\dagger h^\dagger g^\dagger
):_{\tilde m} +
\right. \\ &&
 \left. {1\over 2} {(\partial _- A_+)}^2 -eA_+ (J^+ +j^+_{ext}) \right ],
 \eer
where $tr$ is over $U(N_f\times N_c)$, $g\in SU(N_f)$, $h\in SU(N_c)$, $\exp
 ({-i\sqrt {{4\pi}\over
    {N_cN_f}}}\phi) \in U(1)$,  $l\in U(N_c \times N_f) / SU(N_c)
\times SU(N_f) \times U(1)$, $S^{WZW}_{(k)}$ is the level k WZW
action. $:\ \ :_{\tilde m}$ denoted normal ordering at mass scale
$\tilde m$ and  $m^2=m_q\tilde m C$ where  $m_q$ is the quark mass,
 and $C={1\over2}e^\gamma$ with $\gamma$ Euler's constant.


The bosonized  mass term is the only term in the action that couples
the colored and flavored sectors.
Since we are interested only in the form of the potential we can simplify
the analysis of the equations of motion
of the full theory by restricting
  ourself only to the
colored sector. This can be achieved by setting
$g=1$, $l=1$ and $\phi=0$. The normal ordering  can be performed
 first  at the scale $\mu={e\sqrt{N_f}\over
\sqrt{\pi}}$ and then one can
replace the two mass scales $\mu$ and $m_q$
by a single scale by normal ordering at a certain scale
$m$\cite{FS}. In that case $m=[N_fm_q C \mu^{1-\Delta_c}]^
{1\over 2-\Delta_c}$ where $\Delta_c$  the dimension of $h$ is
given by $\Delta_c={N_c^2-1\over N_c(N_c+N_f)}$

 which leads the following action

\beq
 S= S^{WZW}_{(N_f)} (h)+
 \int d^2x \ tr \left [m^2 (h + h^{-1})+
 {1\over 2} {(\partial _- A_+)}^2 -eA_+ (J^+ +j^+_{ext}) \right ],
 \eeq

and the trace is over the color degrees of freedom.
Equations  \eqref{seta} now read
\bea \label{setab}
 && \partial _+ J^+  +e[A_+,J^+]= {eN_f \over 2\pi}\partial _- A_+ -
 im^2(h^{-1}-h) \\
 && -\partial _-^{\ 2}  A_+ =e(J^+ + j^+_{ext})
\eea

The difference from the equations derived in the massless case is that
here there is a dependence on $h$ on top of the dependence on the currents.
Unlike the abelian theory here we can write a closed expression for the
group element ($h$) only in a formal form in terms of the current.
However we can express it
as an expansion in powers of $J^+$.
Recall tht the current $J^+$ is
related to the color group element $h$ by
\beq
 \label{current} J^+={iN_f\over 2\pi} h\partial _- h^{-1}
\eeq
The expanding   of the inverse relation  takes the form
\bea
 && h=1-{2\pi \over iN_f}{1\over \partial_-}J^+ +{({2\pi \over
     iN_f})}^2({({1\over \partial _-}J^+)}^2-{1\over \partial
   _-}({1\over \partial _-}J^+)J^+)+ \ldots \\
 && h^{-1}=1+{2\pi \over iN_f}{1\over \partial_-}J^+ +{({2\pi \over
     iN_f})}^2{1\over \partial _-}(( {1\over \partial _-}J^+)J^+)+ \ldots
\eea
where the dots denote additional terms which are higher powers of $J^+$.
This expansion makes sense upon the substitution of
$J^+=\sum_{n=0}^\infty {({1\over N_f})}^n J^{(2n)}$ given in
eqn.\eqref{expansion}. Moreover, note that even for the free
level $N_f$ WZW model
this expansion is justified since $J^+$  behaves like $J^+\sim
\sqrt{ N_f}$ as can be deduced from the associated  affine Lie
algebra.
For the expression  needed in \eqref{setab} we get
\beq
 h^{-1}-h = 2{2\pi \over iN_f} {1\over \partial_-} J^+ +
  {({2\pi \over iN_f})}^2  {1\over \partial_-} [{1\over \partial_-}
  J^+,J^+]  + \ldots,
\eeq
Thus, the set of equations \eqref{set}, for static solutions, take
the following form in the presence of mass
\bea
 && \half \partial_1 J^+ + e[A_+,J^+] \simeq -{eN_f\over 2\pi}\half \partial_1
 A_+ \nonumber \\
&& \,\,\,\,\,\,\,\,\,\,\,\,\,\,\,\,\,\,\,
- im^2  ( -2{\sqrt 2 2\pi \over iN_f} {1\over \partial_1} J^+ +
  {({\sqrt 2 2\pi \over iN_f})}^2  {1\over \partial_1} [{1\over \partial_1}
  J^+,J^+] ) \nonumber \\
 && {1\over 2} \partial_1^2 A_+ = -e(J^+ +\half \rho +\half j) \nonumber \\
 && \partial_1j +\half e [A_+,\rho]+\half e[A_+,j]=0 \nonumber
\eea
which to  leading order, in ${1\over N_f}$,  is
\bea
 && \half \partial_1 J^+  = -{eN_f\over 2\pi}\half \partial_1
 A_+
+im^2  2{\sqrt 2 2\pi \over iN_f} {1\over \partial_1} J^+ \nonumber \\
&& {1\over 2} \partial_1^2 A_+ = -e(J^+ +\half \rho) \nonumber
\eea
 Eliminating $J^+$, we  find the following equation for $A_+$
\beq
 (1+{e^2 N_f^2 \over 8\pi^2 m^2}) \partial_1 ^2 A_+ \simeq -e\sqrt 2 \rho
\eeq
The solution of the equation in the presence of \eqref{rho2} is
\beq
A_+ = -{e\over \sqrt 2}(1+ {e ^2 N_f^2\over 8\pi ^2 m^2})^{-1}(Q \mid x-R
\mid +\bar Q \mid x+R \mid)
\eeq
Substituting $A_+$ in the potential yields,
\beq
 V=- {e^2\over 2} (1+ {{e^2 N_f^2}\over{8\pi^2 m^2}})^{-1}Q\bar Q \times 2R
\eeq
Using the definition $d\equiv 2R$, substituting $m^2$,  $\mu ^2$
and
 $Q\bar Q$ which  is minus the quadratic
casimir operator of the test charges in a general representation
${\cal R}$ we obtain
\beq
\label{tension}
 V= {\mu^2 \pi \over 2 N_f} (1+ {\mu \over 8\pi Cm_q} )^{-1}C_2({\cal R})
\times d \eeq
The same expression for the potential in the abelian case was
obtained in \cite{gross}.
Thus the dominant ${1\over N_f}$ contribution exhibits a
confinement behavior.
It should be emphasized that in the above analysis it is assumed
that the external charges cannot be composed by the dynamical ones.
This is the analog of  the abelian case where the
external charges are not integer multiple of the dynamical charges\cite{CJS}.
It is well known  that in systems, where the latter does not hold,
the string between the external quark and antiquark   can be
torn apart by a pair creation.
Similarly in the
non-abelian case we expect that when the test charges can be composed
of a multiplication of the dynamical charges the
string could be  torn. Confinement is restored (and \eqref{tension}
is valid) only when when the test charges {\em cannot be   composed}
of the dynamical charges\cite{witten}.
It is clear that in the latter case the
confining nature of the theory  survives
 higher order
 $1\over {N_f}$ corrections.

\section{Supersymmetric Yang-Mills}
Next, we would like to 
 investigate the large distance behavior of
$N=1$ two dimensional supersymmetric Yang-Mills  theory,
and test the conjecture that the system is in a screening phase \cite{gross}.

The \sym action \cite{ferrara} is the following
\beq
 \label{sym} S = \int d^2 x \ tr \left ( -{1\over 4} F^2_{\mu\nu} +  i\bar
 \lambda\Dslash\lambda
 +{1\over 2} (D_\mu \phi )^2 + 2ie \phi \bar \lambda \gamma _5 \lambda
\right ) ,
\eeq
where  $A_\mu $  the
 gluon field, $\lambda $ the gluino and $\phi $
 a pseudo-scalar - the components of the vector supermultiplet-
transform in the adjoint representation of $SU(N_c)$.

The action is invariant under the following supersymmetric  transformations
\ber
&&  \delta A_\mu = i \bar \epsilon \gamma _5 \gamma _\mu  \sqrt{2} \lambda \\
&&  \delta \phi = - \bar \epsilon \sqrt{2} \lambda \\
&&  \delta \lambda \ = -{1\over {2 \sqrt{2}}} \epsilon \epsilon ^{\mu \nu}
  F_{\mu \nu} +{i \over {\sqrt{2}}} \gamma ^{\mu} \epsilon D_{\mu} \phi
\eer
where $\epsilon$  is the fermionic  parameter of transformation.
In order to determine the behavior of  the theory  it
seems  natural to
follow the same procedure  used in the previous sections,
namely,
to bosonize the fermionic degrees of freedom.
Then, to proceed by solving the equations of motion for the gauge
fields and deducing the potential between the external sources.
 In the present case
one needs to bosonize the
 gluinos which transform in the $SU(N_c)$
adjoint representation. The bosonized version of adjoint fermions
can be expressed in terms of a $SU(N_c)$  $WZW$ model of level
$N_c$\cite{AGSY}.
The  form of the Yukawa term follows from the identification
of $:h-h^{-1}:$ with $:\bar \lambda \gamma_5\lambda:$. 
The full bosonized action in the $A_-=0$ gauge takes
 the following form
\bea
\label{susyaction} \lefteqn{S_{bosonized} = S^{WZW}_{(N_c)}} \\ &&
 + \int d^2 x \ tr \left( {1\over 2}(\partial_- A_+)^2
+{1\over 2} (D_\mu \phi)^2 +i\tilde m e \phi :(h-h^{-1}):_{\tilde m}
 - e A_+( J^+
+ j^+_{ext})
 \right),
\nonumber
\eea
where $\tilde m$ is the scale at which the normal ordering is
performed in a similar way to the one introduced in eqn.
\eqref{ghlboz}.
Eliminating the scalar and bosonized fermion degrees of freedom from
the equations of motion which arise from\eqref{susyaction}, one finds  that
 the effective equation of motion
of  static gauge field is
 \beq
 \label{susyA}
\left (
\dx ^2 - {{e^2 N_c}\over \pi} { 1 \over {1-(2\tilde m e)^2 {{4\pi}
      \over N_c}{1\over{\dx ^4}}}} \right ) A_+ = -\sqrt 2 e \rho
\eeq
The meaning of this equation is that the  corresponding
 propagator of the gluon
 has three poles and one of them is {\em tachyonic}.
It turns out, as will  shown below, 
that the source of this  unwanted pole is the fact that the
fermionic contribution  includes
quantum corrections in the form of the axial anomaly, whereas
the scalar contribution is a tree level one.
This  follows from the fact that we write down the classical
equations of motion  of \eqref{susyaction}  but
it is well known that the tree level bosonized action
incorporates the  axial anomaly.

The correct procedure for the supsrsymmeric case is to consider
 loop effects of both the fermions and the scalars
on the gluon propagator.

Before we describe the calculation associated with
\sym model, let us review some
properties of the Schwinger model and two dimensional scalar
electrodynamics.

The lagrangian of massless $QED_2$ (the Schwinger model) is
\beq
  {\cal L}= -{1\over 4} F^2_{\mu\nu} +  i\bar \Psi \Dslash\Psi
\eeq

The resulting classical equation of motion for the gauge field is
\beq
 \label{schwinger} \Box F = e \epsilon ^{\mu \nu} \partial _{\mu} j_{\nu} ,
\eeq
where $F={1\over 2} \epsilon ^{\mu \nu} F_{\mu \nu}$ and $j^{\mu}
= \bar \Psi \gamma ^{\mu} \Psi$. Classically, the right hand side of
the above equation vanishes because $\partial _{\mu} \epsilon ^{\mu \nu}
j_{\nu} = -\partial _{\mu} j^{\mu}_5 = 0$. However, quantum
mechanically
 the axial current is anomalous  $\partial _{\mu} j^{\mu} _5 = {e
   \over \pi} F$ and therefore the quantum mechanical form of equation
 \eqref{schwinger} is
\beq
 (\Box + {e^2 \over \pi} ) F = 0
\eeq
 which leads to the screening behavior.

 In scalar electrodynamics similar
phenomenon occurs. The lagrangian of
 the theory
\beq
  {\cal L}= -{1\over 4} F^2_{\mu\nu} + (D_\mu \phi )(D^\mu \phi)^\star
\eeq
yields the following classical equation of motion for  the gauge
field  \beq
 \label{scalar} \Box F = e \epsilon ^{\mu \nu} \partial _{\mu} j_{\nu} ,
\eeq
where the scalar abelian current is
 $j^{\mu} = -i(\phi ^\star \partial ^ {\mu} \phi - \phi \partial
^{\mu} \phi ^\star - 2ie A^{\mu} \phi ^ \star \phi )$.
Note that  unlike in the fermionic sector,  the right hand side of
\eqref{scalar} does not vanish even at the classical level since
there is no conserved
axial current in the scalar  sector. The classical divergence of
the  ``axial current"  is
 \beq \label{axial2}
 \partial _{\mu} \epsilon ^{\mu \nu} j_{\nu} =
 -i\epsilon ^ {\mu \nu} (2 \partial _{\mu} \phi ^\star \partial _ {\nu} \phi
 - 2ie \partial _{\mu} (A_{\nu} \phi ^ \star \phi))
\eeq
This relation is modified quantum mechanically similarly to
the modification in the
Schwinger model.  The one loop vacuum polarization diagram
modifies the axial current non conservation. In scalar
electrodynamics the one loop vacuum polarization is given by the
 following two Feynman diagrams (the dashed line is the scalar
field) \begin{figure}[htb]
  \begin{center}
\mbox{\kern-0.5cm
\epsfig{file=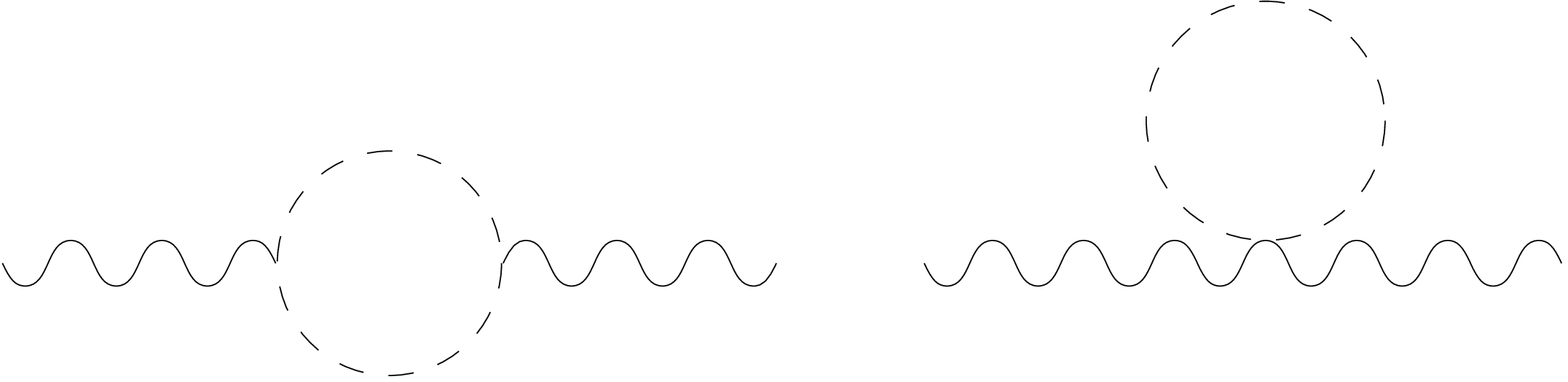,width=10.0true cm,angle=-90}}
\label{vacuum_fig}
  \end{center}
\end{figure}
which combine to
\beq
\Pi _{\mu \nu} (k^2) = (g_{\mu \nu} - {{k_{\mu} k_{\nu}}\over {k^2}} )
\times {e^2 \over \pi} \int ^1 _0 dx {{k^2 x(x-{3\over 2})} \over{k^2 x (1-x) -
 m^2}}
\eeq
 This expression suffers from severe infra-red divergences in the
 $m^2\rightarrow 0$ limit\cite{coleman}.
 This technical obstacle can be overcome
by the use of a point
 splitting of the vector current\cite{schwinger}.
 \bea
  \lefteqn {j^{\mu} = \lim_{\epsilon \rightarrow 0} \,
 \exp (-ie \int^{x+{\epsilon \over 2}}_{x-{\epsilon \over 2}} A_{\mu}(y)
   dy^{\mu}) \times \nonumber } \\ &&
 -i\left (
 \phi ^\star (x+{\epsilon \over  2})  \partial ^ {\mu} \phi (x-{\epsilon \over
 2}) -
\phi  (x-{\epsilon  \over  2})  \partial ^{\mu} \phi ^\star
(x+{\epsilon \over  2}) - \right.  \nonumber \\ &&
\left. ie A ^{\mu} (x+{\epsilon \over  2})  \phi  (x-{\epsilon \over  2})
 \phi ^\star (x+{\epsilon \over  2})-
ie A ^{\mu} (x-{\epsilon \over  2})  \phi  (x-{\epsilon \over  2})
 \phi ^ \star (x+{\epsilon \over  2}) \right ) \nonumber
\eea
 In fact, the point splitting method enables us not only to get rid of the
infra-red divergence but also to
derive the quantum mechanical
modified version of \eqref{axial2}
\beq
 \partial _{\mu} \epsilon ^{\mu \nu} j_{\nu} = -{e \over {2 \pi}} F -
 i\epsilon ^ {\mu \nu} (2 \partial _{\mu} \phi ^\star \partial _ {\nu} \phi
 - 2ie \partial _{\mu} (A_{\nu} \phi ^ \star \phi))
\eeq
which leads to
\beq
(\Box + {e^2 \over {2\pi}}) F = -ie \epsilon ^ {\mu \nu} (2 \partial _{\mu} \phi
 ^\star \partial _ {\nu} \phi
 - 2ie \partial _{\mu} (A_{\nu} \phi ^ \star \phi))
\eeq
Although the right hand side of the above equation does not vanish, it
is clear that the nonlinear terms are interaction terms that can only modify
the photon mass but can not make it massless.
It is thus clear that   the photon of scalar
electrodynamics behaves  in a similar way to
the one of $QED_2$\cite{bengtsson}.

Returning back to the \sym case two  additional complications
 are introduced.
(i) The gauge  interaction is   a nonabelian one
 and (ii) The gluino interacts with the scalars via a Yukawa term.
  Nevertheless,  we can follow the same
procedure  used for the abelian models and
compute the mass of the  gluon pole.
The gluon equation of motion  now reads
\beq
\label{gluon}
D_{\mu} D^{\mu} F = e \epsilon ^{\mu \nu} D_{\mu} (J^\lambda_\nu +
J^\phi _\nu),
\eeq
where $J^\lambda_\mu$ denotes the gluino vector current and $J^\phi
_\mu$ denotes the scalar vector current.
 The equation for  the divergence of the 
 fermionic axial current.
\beq
 \epsilon ^ {\mu \nu} D_\mu J^\lambda _\nu = -{eN_c \over \pi} F + ie
(\phi \bar \lambda \lambda + \bar \lambda \lambda \phi)
\eeq
The factor $N_c$ which appear in the anomaly term is due to
the
 adjoint gluinos which  run in the
anomaly loop. A similar equation holds for the scalar current
\beq
 \epsilon ^{\mu \nu} D_\mu J^\phi _\nu = -{eN_c \over {2\pi}} F -
 \epsilon ^{\mu \nu} \partial _\mu (-2i[\phi ,\partial _\nu \phi] +
 2e[\phi,[A_\nu,\phi]])
\eeq
 Thus the quantum version of equation \eqref{gluon} is
\beq
(D_{\mu} D^{\mu} +{3e^2 N_c \over {2\pi}})F =  ie^2  (\phi \bar
 \lambda \lambda  + \bar \lambda \lambda \phi) -e \epsilon ^{\mu \nu}
 \partial_\mu
(-2i[\phi ,\partial _\nu \phi] + 2e[\phi,[A_\nu,\phi]])
\eeq
 which means that the gluon propagator has only a single pole which is
massive.
 Though our results are based
 on one loop calculations, it seems that higher order corrections -
 which involve gluino and scalar interactions cannot spoil the massive
 nature of the gluon (but may shift its mass).
The implication of the  last equation on the potential between external
charges is clear. The interaction mediated by the exchange of these  massive
modes is necessarily a screening one.
The potential  takes the form of eqn.\eqref{potential2} with a range that
behaves like $\sim [e\sqrt{N_c}]^{-1}$.

\section{Summary}
The study of the interplay between screening and confining phases
is a basic question in strongly interacting systems.
Usually due to the lack of adequate perturbation expansion, exact
statements about confinement versus screening cannot be made.
One system that evades this fate is $QCD_2$. Special powerful
techniques that are applicable only in  2D systems enable one
to derive significant results.
It is because of these results that  $QCD_2$ may serve as an
important laboratory to study real life  $QCD$.

Evidence that massless $QCD_2$, regardless of 
 the quarks representations, is in a screening phase
was presented in the paper of Gross et. al.\cite{gross}.
The dynamical quarks were shown to screen external charges even if
the latter are in a
 representation that  cannot be composed of those of the
dynamical ones.
It was further shown \cite{gross} that once a mass term is  turned
on the dynamical fermions develop a non-vanishing string tension
namely,  a confining potential. Tensionless  strings occur only in
the particular cases that the representation of the external
quarks could be gotten by a composition of the dynamical ones.
These results were argued using  several different methods.
In particular, the potential was extracted by substituting an  
abelian solution of the equations
of motion of both the abelian theory and certain non-abelian ones.

In the present paper we derive additional supporting evidence for
the picture drawn in \cite{gross}: (i) We justify the use of the
equations of motion by applying a large $N_f$ expansion;
(ii) We extend the prove of screening mechanism by using
``semi-classical" external charges;
(iii) Show the confining potential for a bosonized massive model;
(iv) Show that the 2D super YM theory has a screening behavior.

A given massless multi-flavor $QCD_2$ model is a point
in the two dimensional $(N_c,N_f)$ grid. The domain of $N_f=1$
and  large $N_c$ was described in the seminal work of  't Hooft
\cite{tHooft} in the form of the confining mesonic spectrum.
The analysis of 't Hooft was
insensitive to the question of whether the quarks are massless or
massive. One may wrongly get the impression that the massless
theory confines. In fact, this limit is not adequate for the
study
of the question of confinement versus screening.
 Assume that the potential is of the form of leading term of
\eqref{potential}.  Recall that in the large $N_c$ limit,
 $e^2 N_c$ is kept
finite. This implies that the potential
behaves like $(1-e^{-{\tilde\mu\over\sqrt{N_c}}R})\sim
{\tilde\mu\over\sqrt{N_c}}R\times(1-{1\over 2}
{\tilde\mu\over\sqrt{N_c}}R+o({1\over N_c}))$ for fixed $R$ and
large $N_c$, with $\tilde\mu ^2 = {e^2 N_c \over \pi}$
 a finite constant.
Now it is clear that in the limit of
$N_c\rightarrow \infty$ the potential looks like a linear
potential, and thus one cannot  discriminate between the two
scenarios.

The opposite corner in the grid of theories is that of finite
$N_c$ and large $N_f$ with  $e^2N_f$  kept
finite. It can be shown\cite{AdiCobi} that this limit corresponds
to an approximate system of $N_c^2-1$ abelian theories.
In that case the screening nature can be attributed to an exchange of
Schwinger-like massive modes.
The present work, as well  as \cite{FS},  indicates that no
``phase transition" should be expected when passing
to models with small number of flavors.
The screening nature of the potential may indicate
that the theory  with finite $N_c$ has in its spectrum states of masses
of the order of $e$. 
In the large $N_f$,  using the analogy with the massive Schwinger
model, one can get a general picture of the passage to a confining
behavior. The mass of the massive state of the Schwinger model is
shifted once quark mass is turned on. But an additional light
state emerges. Exchange of the latter mode causes confinement.

We would like to emphasize again, that
studying  the  quantum system by  analyzing the
corresponding equations of
motion is a justified approximation only provided that the classical
configurations dominate the functional integral.
This condition is obeyed in the large $N_f$ limit since in that
case an   $N_f$  factor that can be put in front of the whole action
of the colored sector, and thus  plays the role of ${1\over \hbar}$.

Another improvement over the analysis of \cite{gross} is achieved by 
implementing  the
idea of \cite{adler,giles}
to introduce
the density of the external quarks
 in terms of non-commuting matrices.
In that way the non-abelian nature of the large $N_f$ limit
of theory
manifests itself in the form of non vanishing commutator terms
whereas   using the ansatz of \cite{gross} for the
external charges has an abelian behavior in the same limit.

It may seem that the analysis of the  question of screening versus
confinement in $SYM_2$ could follow very similar
lines as those of $QCD_2$.
However, it turns out that the equations of motion
that follow from the action expressed in terms of bosonized gaugino
fields, lead to unacceptable conclusion. In particular it reveals
a tachyonic pole to the gauge field. This situation occurred due to
an unbalanced treatment of the gaugino and scalar degrees of
freedom. It is only after including 
quantum correction also to  the scalar fields, in
the form of point split currents, that a meaningful result
could have been extracted. The latter corresponds
indeed to a screening phase as was conjectured
in \cite{gross}.

Many open questions associated with the interplay between
confinement and screening are still unresolved. One is the
derivation of the string tension for massive dynamical quarks
in any representation of color group and for  any representation of  the
external quarks. 
It is speculated that this string tension
is $\sim \mu m$, where $m$ is the dynamical quark mass and $\mu$ is
 proportional to the coupling constant,
whenever the test charges {\em cannot} be  composed of the
dynamical charges, whereas  when the  test charges {\em can} be 
composed of the dynamical
charges - the string tension is expected to be vanish.
The supersymmetric $YM_2$ has further extensions.
In particular the large $N_c$ limit of the $N=8$ supersymmetric
case has an important significance in relation to the matrix
model representation of M theory.
Last but not least is obviously the
implications to 4D $QCD$. In particular an interesting question
is whether the 2D
large $N_f$ expansion sheds any light on the 4D one.


\section*{Acknowledgements}
We are grateful to S. Yankielowicz for participating in the early stages of
this project and for many useful conversations.
We would like  also to thank  A. Zamolodchikov for illuminating discussions.
One of us (J.S) would like to thank Y. Frishman for many fruitful discussions.
One of us (A.A) would like to thank N. Itzhaki for several illuminating
 discussions.

\newpage
\appendix
\section{Appendix - The $Q \bar Q$ Algebra}

The Appendix is based on refs. \cite{adler,giles}.

Regarding the components of the Gluon fields as operators, ordering
ambiguities may appear in products such as $ f^{abc} \dx A^b_+ A^c_+$.
These  ambiguities are resolved by the transcription
\beq
 [u,v] ^ a = {1\over 2}f^{abc}(u^b v^c + v^c u^b)
\eeq
where $a,b,c,\dots = 1,2,3, N^2-1$ are $SU(N_c)$ indices. The algebra is
spanned by four operators: $Q^a \otimes 1,\ 1\otimes \bar Q^a,\
f^{abc}Q^b \otimes \bar Q^c$ and $d^{abc}Q^b \otimes \bar Q^c$. Where
$Q^a,\ \bar Q^a$ are $SU(N_c)$ generators.

Using the following identifications
\bea
 && e^a_1 = {2\over N}(Q^a \otimes 1 + 1\otimes \bar Q^a),  \\
&& e^a_2 = {4\over N^2}(Q^a \otimes 1 - 1 \otimes \bar Q ^a) -{4\over
  N}d^{abc} Q^b \otimes \bar Q^c,    \\
&& e^a_3 = -{4\over N} f^{abc} Q^b \otimes \bar Q^c,  \\
&& e^a_4= {(N^2-4) \over 4N}(Q^a \otimes 1 - 1 \otimes \bar Q^a) +
d^{abc} Q^b \otimes \bar Q^c
\eea
It was shown that
\bea
  && [e_i,e_j]^a = \epsilon _{ijk} e ^a_k  \; \; \;   i,j,k=1,2,3 \\
  && [e_i,e_4]^a = 0
\eea
 and the set is orthogonal
\beq
 tr \ e^a_i e^b_j = {4\over N}\delta _{ij}\delta ^{ab}
\eeq

This means that the $SU(N_c)$ algebra is actually reduced to a
$SU(2)\otimes U(1)$ problem.

Using the above algebra, it is easy to see that  $tr\ [Q,\bar
Q]Q=tr\ [Q,\bar Q]\bar Q=0$ and $tr\ [Q,\bar Q][Q,\bar Q]= {N^2 \over
  16}tr\ e^a_3 e^a_3 = {N^2\over 2}{N^2-1\over 2N}$.

\newpage
\section{Appendix - Calculation of $A^{(5)}$}
This Appendix contains the calculation of $j^{(4)}$ and $A^{(5)}$
which are needed to determine the potential in the case of massless
dynamical quarks (Section 3).

The next to next to leading order set of equations which determines
$A^{(5)}$ is
\ber
 &&  \half \dx J^{(4)} + [A^{(3)},J^{(0)}] + [A^{(1)},J^{(2)}] =
 -{1\over 2}\mu ^2 \half \dx A^{(5)} \\
  && {1\over 2} \dx ^2 A^{(5)} = -(J^{(4)} + \half j^{(4)}) \\
  &&\dx j^{(4)} + \half [A^{(3)},\rho] +\half [A^{(1)} , j^{(2)}]=0
\eer

After substitution of $A^{(1)}$, $A^{(3)}$ and $J^{(0)}$, $J^{(2)}$ we
find
\ber
 && J^{(4)}= -{1\over 2} \mu ^2 A^{(5)} \\
 && (\dx ^2 -\mu ^2) A^{(5)} = -\sqrt 2 j^{(4)}
\eer
which is very similar to the leading and next to leading order sets of
 equations. The source
 $j^{(4)}$ can be calculated by using the values of $A^{(1)}$ and
 $A^{(3)}$
\beq
  j^{(4)} =
 -\half {1\over \dx} [A^{(3)},\rho] -\half {1\over \dx} [A^{(1)},j^{(2)}]
\eeq
  the result  of the calculation is
\ber
 \lefteqn{ j^{(4)}=} \\ &&
 -{1\over {8 \mu ^3}} e^{-2 \mu R} (1-e^{-2\mu R}) \epsilon (x-R)
 \times [[\bar Q,Q],Q] \\ &&
-{1\over {8 \mu ^3}} e^{-2 \mu R} (1-e^{-2\mu R}) \epsilon (x+R)
 \times [[\bar Q,Q],\bar Q] \\ &&
+{1\over {8 \mu ^3}} \left (\epsilon (x-R) - \epsilon (x+R))(1- e^{-\mu
  \mid x - R \mid}) + (\epsilon (x+R)+1) (1-e^{-2\mu R}) \right ) \times
 [[\bar Q, Q],Q] \\ &&
+{1\over {8 \mu ^3}} \left (\epsilon (x+R) - \epsilon (x-R))(1- e^{-\mu
  \mid x + R \mid}) + (\epsilon (x-R)+1) (1-e^{-2\mu R}) \right ) \times
 [[\bar Q, Q],\bar Q]
\eer
The first two lines in the expression of $j^{(4)}$ arise from the
commutator of $A^{(3)}$ and $\rho$, the other lines are due to the
commutator of $A^{(1)}$ and $j^{(2)}$. Using $j^{(4)}$, a tedious
calculation yields the following
  expression of  $A^{(5)}$
 \ber
 \lefteqn { A^{(5)}=} \\ &&
  -{\sqrt 2 \over {8 \mu ^ 5}} e^{-2 \mu R} (1-e ^{-2 \mu R})
 \epsilon (x-R)(1-e^{- \mu \mid x-R \mid } ) \times [[\bar Q,Q],Q] \\ &&
 -{\sqrt 2 \over {8 \mu ^ 5}} e^{-2 \mu R} (1-e ^{-2 \mu R})
 \epsilon (x+R)(1-e^{- \mu \mid x+R \mid } ) \times [[\bar Q,Q],\bar Q]
 \\ &&
  -{\sqrt 2 \over {8 \mu ^5}}  \times
   \left \{\begin{array}{ll}
          -2\mu R e ^ {\mu (x-R)}                                     & x<-R \\
          2e^{-2 \mu R} +(\mu x-\mu R-{3\over 2}) e^{\mu (x-R)}  &  -R<x<R \\
           - {1\over 2} e^{-4\mu R} e^{-\mu (x-R)}      \\
           -2(1-e^{-2\mu R}) + {1\over 2}(1-e^{-4 \mu R})e^{-\mu
             (x-R)}                & x>R
       \end{array}
   \right \}
   \times [[\bar Q,Q],Q]
  \\ &&
  -{\sqrt 2 \over {8 \mu ^5}}  \times
   \left \{\begin{array}{ll}
            -{1\over 2}(1-e^{-4\mu R}) e^{\mu (x+R)}            & x<-R \\
           -2  +(\mu x+\mu R+{3\over 2}) e^{-\mu (x+R)}  &  -R<x<R \\
           +{1\over 2} e^{-4\mu R} e^{\mu (x+R)}      \\
           -2(1-e^{-2\mu R}) + 2\mu Re^{-\mu(x+R)}             & x>R
       \end{array}
   \right \}
   \times [[\bar Q,Q],\bar Q]
\eer

\end{document}